# Actual and weak actual values in Bohmian mechanics

Weixiang Ye

Center for Theoretical Physics, School of Physics and Optoelectronic Engineering, Hainan University, Haikou 570228, China

wxy@hainan.edu.cn

**Abstract**

We systematically analyze Holland's local expectation values within Bohmian mechanics, referring to them as weak actual values to emphasize their connection with weak measurement theory. We derive the exact time evolution equation for these quantities along a Bohmian trajectory and formally establish their correspondence with the real part of the weak value under position postselection. The explanatory power of this framework is demonstrated by revisiting a recent waveguide experiment: we show that the measured quantity corresponds to the imaginary part of the momentum weak value, which reflects the spatial decay of the wave field, not the particle velocity. This cleanly clarifies the distinct physical roles of the real and imaginary parts of weak values.

## 1 Introduction

The interpretation of quantum mechanics remains an actively debated foundational topic more than a century after its inception [1–4]. Among various interpretive frameworks, Bohmian mechanics, also known as the pilot-wave or de Broglie-Bohm theory, stands out for its explicit ontological commitments and deterministic dynamics [5–8]. In this theory, particles are postulated to possess definite positions at all times, with their motion governed by a guiding equation, while the wave function evolves according to the Schrödinger equation. This framework reproduces all statistical predictions of standard quantum mechanics. A long-standing foundational question concerns the ontological status of observables other than position. In standard quantum mechanics, observables are represented by self-adjoint operators whose values are revealed by measurement. In Bohmian theory, position is taken as the sole fundamental variable [9–11]. Dürr, Goldstein, and Zanghì [12] systematically developed the statistical foundations of Bohmian mechanics, emphasizing that position always has an actual value, while momentum has one only in an asymptotic sense and spin generally does not. Norsen [13] provided a clear physical argument showing that a spin component acquires a definite value only when the wave function is in an eigenstate of the corresponding Pauli matrix. A particularly relevant framework was developed by Holland [8], who introduced the concept of local expectation values for arbitrary observables. Defined as $\mathrm{Re}\,[\psi^*(\hat{A}\psi)]/|\,\psi\,|^2$, these quantities provide a local description along Bohmian trajectories and satisfy important statistical properties, notably that their ensemble average equals the quantum expectation value [8]. In this paper, we refer to these quantities as weak actual values. This choice of terminology is intended to emphasize that the quantity is a legitimate actual (i.e., trajectory-defined) attribute of the Bohmian particle, but one that is intimately related to the weak value of quantum measurement theory. The adjective *weak* modifies the noun phrase *actual value*, distinguishing this derived construct from both the strong (projective) measurement values and the standard operational weak value

Our work builds upon and extends Holland's foundational analysis in three principal ways. First, we derive the exact time evolution equation for the weak actual value along a Bohmian trajectory (Theorem 2), thereby adding a dynamical dimension to what was primarily a static picture. The three terms in this equation admit a clear physical interpretation that we discuss in detail. Second, we formally establish the precise relation between the weak actual value and the real part of the weak value in quantum measurement theory under position postselection (Eq. (15)), generalizing earlier insights to arbitrary observables. From this correspondence we extract three principal lessons that succinctly capture the physical connection between Bohmian mechanics and weak measurements. Third, and most importantly, we demonstrate the explanatory power of this unified framework by analyzing a recent experimental challenge [14–17], clarifying the distinct physical roles of the real and imaginary parts of weak values thereby cleanly resolving an apparent contradiction with standard Bohmian mechanics. Recent work by Foo et al. [18,19] has explored connections between weak values and Bohmian trajectories in relativistic contexts, complementing our nonrelativistic analysis. The structure of this paper is as follows. Section 2 reviews the basic equations of Bohmian mechanics and introduces a heuristic criterion for conceptually exploring the question of definiteness. Section 3 introduces the weak actual value, studies its properties, derives its evolution equation, and establishes its correspondence with weak measurement theory. Section 4 applies the framework to specific observables: position, momentum, spin, and energy, and discusses compatibility with foundational theorems. Section 5 applies the framework to resolve the recent waveguide experiment. Section 6 concludes and outlines directions for future research. Appendices provide detailed proofs and derivations.

# 2 Actual values in Bohmian mechanics: a heuristic criterion

### 2.1 Fundamental equations of Bohmian mechanics and quantum equilibrium

We consider a nonrelativistic quantum system composed of $N$ particles. The complete physical state is jointly described by the wave function $\psi$ and the actual spatial configuration $Q = (\mathbf{Q}_1, \ldots, \mathbf{Q}_N)$, where $\mathbf{Q}_k \in \mathbb{R}^3$ is the position of the $k$-th particle. The wave function $\psi(\mathbf{q}, t)$, defined on configuration space $\mathbb{R}^{3N}$, evolves according to the Schrödinger equation

$$i\hbar \frac{\partial \psi(\mathbf{q}, t)}{\partial t} = \hat{H}\psi(\mathbf{q}, t), (1)$$

where the Hamiltonian takes the nonrelativistic form

$$\hat{H} = -\sum_{k=1}^{N} \frac{\hbar^2}{2m_k} \nabla_k^2 + V(\mathbf{q}). (2)$$

Here, $m_k$ is the mass of the $k$-th particle, $V(\mathbf{q})$ is the potential energy function, and $\nabla_k^2$ denotes the Laplacian with respect to $\mathbf{q}_k \in \mathbb{R}^3$.

Simultaneously, each particle possesses a definite position $\mathbf{Q}_k(t)$, and its motion is described by the Bohmian guiding equation [11,20]

$$\frac{d\mathbf{Q}_k(t)}{dt} = \mathbf{v}_k^{\psi}(Q(t), t) = \frac{\hbar}{m_k} \mathrm{Im} \left( \frac{\psi^* \nabla_k \psi}{\psi^* \psi} \right) |_{\mathbf{q}=Q(t)}. (3)$$

For multi-component spinor wave functions, $\psi^* \nabla_k \psi$ is understood as the scalar product in spin space. Equations (1) and (3) together constitute a deterministic system: given the initial wave function $\psi_0$ and the initial configuration $Q_0$, the future evolution is uniquely determined.

To recover the statistical predictions of standard quantum mechanics, the quantum equilibrium hypothesis is required. This hypothesis stipulates that the probability density of the initial configuration $Q_0$ is given by the squared modulus of the initial wave function [9,11,21]

$$\rho_0(\mathbf{q}) = | \psi_0(\mathbf{q}) |^2. (4)$$

Under this hypothesis, it can be proven that the configuration distribution at any time $t$ satisfies $\rho_t(\mathbf{q}) = | \psi_t(\mathbf{q}) |^2$. This property, called equivariance, guarantees that the statistical predictions of Bohmian mechanics agree completely with the Born rule of standard quantum mechanics (See Appendix A for a detailed derivation of equivariance).

### 2.2 Conceptual prelude: the local eigencondition

Within the explicit primitive ontology of Bohmian mechanics, where only particle positions $Q$ are fundamental, the status of any other observable $A$ must be analyzed rather than assumed. The natural way to probe this question is to examine whether the wave function, locally at the particle's position, behaves as an eigenfunction of the corresponding operator. This observation leads to the following heuristic starting point.

**Remark 2.1 (A heuristic starting point: the local eigencondition).** Let $\hat{A}$ be a self-adjoint operator representing an observable $A$, acting on the system's Hilbert space $\mathcal{H}$. A Bohmian state is described by the wave function $\psi \in \mathcal{H}$ and the particle configuration $Q \in \mathbb{R}^{3N}$. A sufficient condition for the system to possess a definite, time-independent value of the observable $A$ along the trajectory is that the wave function satisfy the local eigencondition:

$$(\hat{A}\psi)(\mathbf{q})\,|_{\mathbf{q}=Q} = \lambda\psi(Q), (5)$$

for some constant $\lambda \in \mathbb{R}$. In such a special case, the constant $\lambda$ can be meaningfully associated with the system along the trajectory and would be termed an actual value of $A$.

**Remark 2.2.** This heuristic criterion has the following physical implications. First, position, as the fundamental variable, always satisfies the local eigencondition (with $\hat{A} = \hat{\mathbf{X}}_k$ and $\lambda = \mathbf{Q}_k$). Second, for non-eigenstates, most observables such as momentum, spin, angular momentum, and energy do not satisfy the local eigencondition. For example, the momentum operator $\hat{\mathbf{P}}_k = -i\hbar\nabla_k$ has a constant associated value only when the wave function satisfies the local eigencondition at the particle's position, which is generally not the case. This captures the insights of Dürr, Goldstein, and Zanghì [12] regarding momentum, and of Norsen [13] regarding spin. The heuristic criterion of the local eigencondition, while conceptually appealing, is too restrictive to be of general use: for the vast majority of quantum states and observables, it simply fails. Nevertheless, it serves an important purpose. It raises the question: if an observable does not satisfy the local eigencondition, is there nonetheless some derived quantity, determined entirely by the wave function and the trajectory, that characterizes the local manifestation of that observable? The affirmative answer to this question is provided by the notion of the weak actual value, which we now introduce.

# 3 Weak actual values: definition, evolution, and correspondence with weak measurements

## 3.1 Definition and basic properties

To characterize the local behavior of observables along a particle trajectory when the local eigencondition does not hold, we now study the weak actual value as a derived theoretical construct. Crucially, this quantity is not a new ontological element but is completely determined by the wave function $\psi$ and the configuration $Q$, serving as a descriptive tool.

**Remark 3.1.** The quantity defined below is mathematically identical to the local expectation value extensively studied by Holland in Ref. [8]. The present work extends Holland's analysis in three directions: we provide an explicit evolution equation for these quantities (Section 3.2), we establish their precise relation to weak measurement theory (Section 3.3), and we deploy this unified framework to resolve a recent experimental challenge (Section 5). Following the insights of Leavens [20], Wiseman [22], and Matzkin [23] on the connection between Bohmian mechanics and weak measurements, we adopt the terminology *weak actual value*. Recent work by Foo et al. [18,19] has explored similar connections in relativistic settings.

**Definition 3.1 (Weak actual value).** For an observable $A$ represented by a self-adjoint operator $\hat{A}$, and a Bohmian state $(\psi, Q(t))$ at time $t$, its weak actual value $a_w(t)$ is defined as a quantity derived from the wave function and the configuration according to

$$a_w(t) \equiv \frac{\mathrm{Re}\left[\psi^*(\mathbf{q},t)(\hat{A}\psi)(\mathbf{q},t)\right]}{|\psi(\mathbf{q},t)|^2}\Big|_{\mathbf{q}=Q(t)}. \quad (6)$$

For spinor wave functions, $\psi^*$ is understood as the Hermitian conjugate $\psi^\dagger$. For differential operators such as momentum, the expression $(\hat{A}\psi)(\mathbf{q},t)$ is understood in the sense of distributions. Under the quantum equilibrium hypothesis, the definition is well-defined for almost all points along a trajectory, as the probability of the particle being at a node ($|\psi|^2 = 0$) is zero.

From the definition, the following basic properties are immediate:

1. **Real-valued:** $a_w(t) \in \mathbb{R}$.
2. **Compatibility with eigenstates:** If the wave function satisfies the local eigencondition $(\hat{A}\psi)(Q(t),t) = \lambda(\psi, Q(t))$ at the particle position, then substituting into (6) yields $a_w(t) = \lambda$. In this special case, the weak actual value reduces to the heuristic constant from Remark 2.1.
3. **Context dependence:** $a_w(t)$ depends on the wave function $\psi$ and the particle position $Q(t)$, reflecting the holistic and context-dependent nature of observable manifestations in Bohmian mechanics.

The key characteristic of the weak actual value lies in its ensemble statistical behavior, which establishes a direct link with standard quantum mechanics.

**Property 1 (Ensemble average).** As shown by Holland [8], for a system in a quantum equilibrium ensemble described by a normalized wave function $\psi$, the ensemble average of the weak actual value equals the quantum expectation value of the observable $\hat{A}$:

$$\mathbb{E}[a_w] = \int \mid \psi(\mathbf{q}) \mid^2 a_w(\mathbf{q})\, d\mathbf{q} = \langle \psi \mid \hat{A} \mid \psi \rangle, (7)$$

where $a_w(\mathbf{q})$ is the configuration-space function obtained by replacing $Q(t)$ with the configuration variable $\mathbf{q}$ in definition (6).

**Proof.** Direct calculation:

$$\mathbb{E}[a_w] = \int \mid \psi \mid^2 \frac{\mathrm{Re}\,[\psi^* \hat{A}\psi]}{\mid \psi \mid^2}\, d\mathbf{q}$$

$$= \int \mathrm{Re}\,[\psi^* \hat{A}\psi]\, d\mathbf{q} = \frac{1}{2}\int \left[\psi^* \hat{A}\psi + (\hat{A}\psi)^* \psi\right] d\mathbf{q}.$$

Because $\hat{A}$ is self-adjoint, $\int (\hat{A}\psi)^* \psi\, d\mathbf{q} = \langle \psi \mid \hat{A} \mid \psi \rangle^* = \langle \psi \mid \hat{A} \mid \psi \rangle$. Hence the right-hand side equals $\langle \psi \mid \hat{A} \mid \psi \rangle$. □

Equation (7) shows that, although the weak actual value $a_w(t)$ varies in space and time along a single trajectory, its statistical average over the quantum equilibrium ensemble precisely reproduces the statistical prediction of standard quantum mechanics. This provides a mathematical foundation for establishing a strict correspondence between the trajectory description of Bohmian mechanics and quantum ensemble statistics.

### 3.2 Time evolution equation

We consider an operator $\hat{A}$ that does not explicitly depend on time. To study the dynamics of the weak actual value along a Bohmian trajectory, we treat it as a function of configuration space and time

$$a_w(\mathbf{q}, t) \equiv \frac{\mathrm{Re}\left[\psi^*(\mathbf{q}, t)(\hat{A}\psi)(\mathbf{q}, t)\right]}{\mid \psi(\mathbf{q}, t) \mid^2}. (8)$$

The weak actual value on the trajectory is then $a_w(t) = a_w(Q(t), t)$. Its total time derivative is given by

$$\frac{da_w}{dt} = \frac{\partial a_w}{\partial t} \mid_{\mathbf{q}=Q(t)} + \sum_{k=1}^{N} \mathbf{v}_k \cdot \nabla_k a_w \mid_{\mathbf{q}=Q(t)}, (9)$$

where $\mathbf{v}_k = d\mathbf{Q}_k/dt$ is the Bohmian velocity given by Eq. (3), and $\nabla_k$ denotes the gradient with respect to the $k$-th particle coordinate $\mathbf{q}_k$.

**Theorem 2 (Evolution equation for the weak actual value).** Let $\hat{A}$ be a self-adjoint operator not explicitly time-dependent. Assume that the wave function $\psi(\mathbf{q}, t)$ is sufficiently smooth such that $\psi$ and $\hat{A}\psi$ are continuously differentiable to second order in $\mathbf{q}$ and once in $t$ in a neighborhood of the trajectory. For initial wave functions that are $C^\infty$-vectors of the Hamiltonian, standard results on the global existence and uniqueness of Bohmian trajectories, specifically the theorem of Teufel and Tumulka [24], guarantee that for almost all initial configurations relative to the $|\psi_0|^2$ distribution, the trajectory $Q(t)$ exists globally in time and never encounters a node of the wave function. Hence, for all physically relevant trajectories (i.e., with probability one), the condition $\psi(Q(t),t) \neq 0$ holds for all $t$. Consequently, the premise that trajectories avoid nodes is not an assumption but a consequence of the general theory. The weak actual value then evolves according to

$$\begin{aligned}\frac{da_w}{dt} = \frac{1}{\rho(\mathbf{q},t)} \mathrm{Re} & \left[ \frac{1}{i\hbar} \left( \psi^* \hat{A}\hat{H}\psi - (\hat{H}\psi)^* \hat{A}\psi \right) \right. \\ & \left. + \sum_{k=1}^{N} \mathbf{v}_k \cdot \nabla_k (\psi^* \hat{A}\psi) \right] |_{\mathbf{q}=Q(t)} \\ & + a_w \sum_{k=1}^{N} \nabla_k \cdot \mathbf{v}_k \, |_{\mathbf{q}=Q(t)},\end{aligned} \tag{10}$$

where $\rho = |\psi|^2$ is the probability density. The evolution equation holds pointwise where $\rho \neq 0$. For wave functions with lower regularity, it can be interpreted in the distributional sense.

**Proof.** The detailed derivation is given in Appendix B.1. □

Equation (10) decomposes the evolution of $a_w$ into three physically distinct contributions:

i. **Quantum dynamical term:** The expression $\frac{1}{i\hbar}(\psi^* \hat{A}\hat{H}\psi - (\hat{H}\psi)^* \hat{A}\psi)$ is the local, configuration-space integrand of the Heisenberg equation of motion for $\hat{A}$. For an operator commuting with the Hamiltonian, $[\hat{A},\hat{H}] = 0$, this entire term vanishes identically. Conversely, when $\hat{A}$ fails to commute with $\hat{H}$, this term drives the evolution of $a_w$ through the inherent quantum dynamics, independent of the particle's spatial motion.
ii. **Convective transport term:** $\sum_k \mathbf{v}_k \cdot \nabla_k (\psi^* \hat{A}\psi)$ captures how the value of $a_w$ changes as the particle is carried by the Bohmian flow through regions where $\psi^* \hat{A}\psi$ varies spatially. Even if the quantum dynamical term vanished, the weak actual value could still evolve due to the particle traversing regions with different local expectation properties.
iii. **Probability flow correction term:** $a_w \sum_k \nabla_k \cdot \mathbf{v}_k$ arises from the divergence of the Bohmian velocity field. This term maintains consistency with the evolution of the probability density $\rho$: as the ensemble of trajectories converges toward or diverges from a region, the density $\rho$ changes on the ensemble level, and this correction ensures that the statistical relationship expressed by Property 1 is preserved at every instant of time.

**Corollary 1.** For a time-dependent operator $\hat{A}(t)$, the evolution equation becomes

$$\frac{da_w}{dt} = \text{RHS of Eq. (10)} + \frac{1}{\rho(\mathbf{q},t)}\text{Re}\left[\psi^*(\mathbf{q},t)\frac{\partial \hat{A}(t)}{\partial t}\psi(\mathbf{q},t)\right]|_{\mathbf{q}=Q(t)}. \quad (11)$$

### 3.3 Correspondence with weak measurement theory

The definition of the weak actual value has a profound mathematical connection with the concept of the weak value in quantum weak measurement theory [25,26]. In the standard weak measurement protocol, one performs a weak measurement of an observable $\hat{A}$ on an ensemble of identically prepared systems, each initially in state $|\psi_i(t_i)\rangle$. The systems then evolve freely until a final postselection measurement is made, projecting each system onto a desired final state $|\phi_f(t_f)\rangle$. For the subensemble of systems that are successfully postselected, the average shift of the pointer of the weak measurement apparatus is proportional to the real part of the weak value. The weak value itself is a complex number defined as [25]

$$A_w = \frac{\langle\phi_f(t_f)\mid\hat{U}(t_f,t_w)\hat{A}\hat{U}(t_w,t_i)\mid\psi_i(t_i)\rangle}{\langle\phi_f(t_f)\mid\hat{U}(t_f,t_i)\mid\psi_i(t_i)\rangle}, \quad (12)$$

where $t_w$ is the time of the weak interaction, and $\hat{U}(t_2,t_1)$ is the time evolution operator. If the weak measurement is impulsive and occurs immediately before the postselection (i.e., $t_w = t$ and $t_f = t^+$), and if the free evolution during the infinitesimal interval is negligible, Eq. (12) simplifies to

$$A_w(t) = \frac{\langle\phi_f(t)\mid\hat{A}\mid\psi_i(t)\rangle}{\langle\phi_f(t)\mid\psi_i(t)\rangle}. \quad (13)$$

It is instructive to see how the real part of the weak value emerges from a deterministic trajectory picture. Suppose one assumes, as in Bohmian mechanics, that the particle has an actual position at all times. Then a natural postselection is simply the particle's actual position at the time of the weak measurement. That is, one postselects on the configuration $|\phi_f(t)\rangle = |Q(t)\rangle$. Substituting this into the simplified weak value expression (13) yields

$$A_w(Q(t)) = \frac{\langle Q(t)\mid\hat{A}\mid\psi(t)\rangle}{\langle Q(t)\mid\psi(t)\rangle} = \frac{(\hat{A}\psi)(Q(t),t)}{\psi(Q(t),t)}, \quad (14)$$

where for differential operators such as momentum, $(\hat{A}\psi)(Q)$ is understood as the distributional evaluation of $\hat{A}\psi$ at the point $Q$. This is well-defined under the smoothness assumptions discussed in Theorem 2.

Comparing this with the definition of the weak actual value (6), we immediately obtain the fundamental relation

$$a_w(t) = \text{Re}\,[A_w(Q(t))]. \quad (15)$$

**From this correspondence we learn three principal lessons.**

i. **Ontological grounding of the weak value's real part:** Equation (15) shows that the real part of the weak value, evaluated at the particle's position, coincides with the weak actual value. The latter is a quantity entirely determined by the wave function and the actual Bohmian trajectory. Thus, in Bohmian mechanics, the real part of the weak value acquires a clear ontological status as a local, trajectory-dependent property of the system, not merely an operational pointer shift.

ii. **Operational accessibility of Bohmian trajectories:** Conversely, the correspondence means that weak measurements postselected on particle positions probe, on average, the weak actual values along the Bohmian trajectories. This provides an operational route to empirically access certain aspects of the Bohmian flow via weak measurement experiments, without requiring strong measurements that would collapse the wave function [20,22,27].

iii. **Fundamental distinction between real and imaginary parts:** The weak value is generally complex. Equation (15) isolates the physical meaning of its real part in terms of Bohmian particle kinematics. The imaginary part, as we shall demonstrate explicitly in Section 5, encodes geometric properties of the guiding wave field, such as its amplitude gradients and spatial decay rates. This distinction is not merely formal; it provides a sharp criterion for determining whether a given experimental procedure probes the particle motion or the wave field geometry.

**Remark 3.2.** It is important to distinguish the underlying approaches. Wiseman's derivation [22] starts from the operational framework of weak measurements and shows how Bohmian trajectories can be reconstructed from weak values, thereby providing an operational grounding for Bohmian mechanics. In contrast, our work begins with the established postulates of Bohmian mechanics and defines the weak actual value as a quantity derived from the fundamental variables $\psi$ and $Q$; we then prove its mathematical equivalence to the real part of the weak value under position postselection. This connection was first pointed out by Leavens [20] for the momentum operator, and later explored in a more general context by Matzkin [23] for trajectory reconstruction.

# 4 Physical interpretation and discussion of applications

### 4.1 Position and momentum

Position is the fundamental variable in Bohmian mechanics. For the position operator $\widehat{\mathbf{X}}_k$, which acts as multiplication by $\mathbf{q}_k$ in the position representation, its weak actual value is given directly by the particle position

$$\mathbf{x}_{w,k}(t) = \mathbf{Q}_k(t). \quad (16)$$

This function does not depend on the wave function $\psi$ and obviously satisfies the continuity requirement. This is consistent with the status of position as the primitive ontology of Bohmian mechanics.

The case of momentum is different. According to Remark 2.1, a constant weak actual value for the momentum operator $\widehat{\mathbf{P}}_k = -i\hbar\nabla_k$ exists only when the wave function satisfies the local eigencondition at the particle position, which is generally not the case. However, its weak actual value carries clear physical significance. For a single-particle system, the momentum weak actual value is

$$\mathbf{p}_w(t) = \frac{\mathrm{Re}[\psi^*(\mathbf{q},t)(-i\hbar\nabla\psi(\mathbf{q},t))]}{|\psi(\mathbf{q},t)|^2}\Big|_{\mathbf{q}=Q(t)}. \quad (17)$$

Writing the wave function in polar form $\psi = Re^{iS/\hbar}$ where $R$ and $S$ are real functions, a straightforward calculation yields (see Appendix B.2):

$$\mathbf{p}_w(t) = \nabla S(Q(t),t) = m\mathbf{v}(Q(t),t). \quad (18)$$

This is precisely the momentum appearing in the Bohmian guiding equation (3), which directly guides the particle's motion. From Property 1, the ensemble average $\mathbb{E}[\mathbf{p}_w] = \langle\psi \mid \widehat{\mathbf{P}} \mid \psi\rangle$.

### 4.2 Spin

The status of spin in Bohmian mechanics has been clearly articulated by Norsen [13]: spin components generally lack definite values unless the system is in an eigenstate. Our heuristic criterion from Remark 2.1 provides a formal underpinning: a spin component $\hat{S}_n$ has a constant weak actual value if and only if the wave function satisfies the local eigencondition for the corresponding Pauli matrix at $Q$. In a Stern–Gerlach experiment, spatial separation ensures that within each output beam, the conditional wave function effectively satisfies this local condition, thus creating a definite spin value upon measurement [28]. Our framework recovers this picture through Remark 2.1 and further allows us to compute a weak actual value $S_{n,w}(t)$ for spins in superposition.

For a spin-$\frac{1}{2}$ particle, the spin component operator along direction $\mathbf{n}$ is $\hat{S}_n = (\hbar/2)\boldsymbol{\sigma} \cdot \mathbf{n}$. For superposition states without spatial separation, spin does not satisfy the local eigencondition, but its weak actual value can be calculated. For the spin $z$-component

$$S_{z,w}(t) = \frac{\mathrm{Re}\left[\psi^\dagger(\mathbf{q},t)(\frac{\hbar}{2}\sigma_z)\psi(\mathbf{q},t)\right]}{\psi^\dagger(\mathbf{q},t)\psi(\mathbf{q},t)} |_{\mathbf{q}=Q(t)}. (19)$$

$S_{z,w}$ can take continuous values between $-\hbar/2$ and $+\hbar/2$, and for certain relative phases may lie outside the eigenvalue range, corresponding to the anomalous weak value phenomenon [23,29].

**Remark 4.1.** For a spin-$\frac{1}{2}$ particle described by the Pauli equation, the probability density is $\rho = \psi^\dagger\psi$ and the probability current is $\mathbf{j} = (\hbar/m)\mathrm{Im}\,(\psi^\dagger\nabla\psi)$. A straightforward calculation shows that the momentum weak actual value is simply

$$\mathrm{p}_w = m\mathrm{v}, (20)$$

where $\mathrm{v} = \mathbf{j}/\rho$ is the Bohmian velocity. This identity holds for both scalar and spinor wave functions. For a spinor particle, the Bohmian velocity receives an additional spin-dependent contribution from the imaginary part of $\psi^\dagger\nabla\psi$, which is precisely captured by the weak actual value. Any osmotic-like contribution from the amplitude gradient, which appears in the decomposition of the kinetic term, is purely imaginary and thus cancels when the real part is taken.

### 4.3 Energy

Energy is described by the Hamiltonian operator $\widehat{H}$. According to Remark 2.1, energy has a constant weak actual value only in an energy eigenstate. Its general weak actual value is

$$E_w(t) = \frac{\mathrm{Re}\left[\psi^*(\mathbf{q},t)(\widehat{H}\psi)(\mathbf{q},t)\right]}{|\,\psi(\mathbf{q},t)\,|^2} |_{\mathbf{q}=Q(t)}. (21)$$

For a single particle in a potential $V(\mathbf{x})$ with $\widehat{H} = -\hbar^2\nabla^2/2m + V(\mathbf{x})$, using $\psi = Re^{iS/\hbar}$ yields (see Appendix B.3)

$$E_w(t) = \left(\frac{(\nabla S)^2}{2m} + V(\mathbf{x}) - \frac{\hbar^2}{2m}\frac{\nabla^2 R}{R}\right) |_{\mathbf{x}=Q(t)}. (22)$$

This coincides with the Bohmian energy defined as the sum of kinetic, classical potential, and quantum potential energies. Its ensemble average satisfies $\mathbb{E}[E_w] = \langle\widehat{H}\rangle$.

### 4.4 Consistency with quantum measurement theory and foundational theorems

Our framework offers a coherent picture for understanding measurement processes from weak to strong within the Bohmian context.

i. **Weak measurement:** The average pointer shift for a postselected subensemble is proportional to $\mathrm{Re}\,[A_w]$. According to Eq. (15), this equals the weak actual value $a_w$ for the corresponding postselected position. Weak measurement experiments reconstructing average trajectories [27] can thus be interpreted as probing ensemble properties of the weak actual value. Foo et al. [18,19] extended this connection to relativistic and curved spacetime settings.
ii. **Strong (projective) measurement:** The strong coupling splits the total wave function into macroscopically non-overlapping branches. Once the Bohmian configuration enters the support of a particular branch, the conditional wave function locally satisfies the local eigencondition, and the observable acquires a constant weak actual value (an eigenvalue) in that branch.

**Kochen–Specker theorem and context-dependence.** The Kochen–Specker theorem [30,31] states that non-contextual definite values cannot be assigned to all quantum observables simultaneously. Bohmian mechanics, through the weak actual value, does not assign such values. The assignment is inherently context-dependent, relying on the particle configuration $Q$ and the global wave function, thus elegantly avoiding the paradox [1].

**Bell's theorem and nonlocality.** Bell's theorem [2] demonstrates that any theory reproducing quantum statistics must be nonlocal. Bohmian mechanics is explicitly nonlocal [32]. Our framework is compatible with this: most observables lack constant weak actual values independent of context; when a constant value emerges through measurement, the process reflects this inherent nonlocality.

**Relation to earlier work.** This framework builds upon and unifies prior research. Holland's local expectation values [8] provide the mathematical basis. Leavens [20] first pointed out the connection between weak measurements and Bohmian velocities. Wiseman [22] gave an operational derivation of the Bohmian velocity from weak values. Norsen and Struyve [28] linked weak measurements to the conditional wave function. Matzkin [23] explored the relationship between weak measurements and trajectories. Our heuristic criterion from Remark 2.1 encapsulates the physical insights of Dürr, Goldstein, and Zanghì [12] and Norsen [13], while the evolution equation (10) extends Holland's analysis by providing an explicit dynamical description. Recent work by Foo et al. [18,19] has extended these ideas to relativistic and curved spacetime contexts.

# 5 Revesting the waveguide experiment challenge

A recent debate [14–17,33] concerning an apparent experimental challenge to the Bohmian velocity formula in coupled waveguide systems provides a striking illustration of the explanatory power of the weak actual value framework, and in particular of the third lesson articulated in Section 3.3: the fundamental distinction between the real and imaginary parts of the weak value. The experiment by Sharoglazova et al. [14] studies scattering at a potential step using a photonic waveguide platform. By analyzing the spatial growth of the photon population in an auxiliary waveguide, the authors extract a scale parameter $v$ that satisfies

$$v=\sqrt{\frac{2\mid\Delta\mid}{m}},\Delta=E-V_0+\hbar J_0,(23)$$

for both classically allowed ($\Delta>0$) and forbidden ($\Delta<0$) regimes. For the evanescent regime ($\Delta<0$), this yields a finite, energy-dependent $v$, which has been interpreted by the experimental authors as a "particle speed" in the forbidden region. Since the standard Bohmian prescription predicts zero particle velocity for a stationary, real evanescent wave ($\nabla S=0$), the result appears to contradict the theory.

To resolve this, we must first be precise about what the term "particle speed" operationally means. In any physical theory that postulates particle trajectories, the instantaneous speed of a particle at time t is defined as the magnitude of the time derivative of its position, $\mid\dot{x}(t)\mid$. In Bohmian mechanics, this derivative is given by the guiding equation (3), which is fully determined by the phase gradient of the wave function. Any quantity that is not measured by tracking the temporal displacement of particle positions cannot, strictly speaking, be identified as a particle speed; at best, it is a parameter with dimensions of velocity that characterizes some other aspect of the physical situation. The parameter $v$ extracted in the experiment is obtained by fitting the steady-state relative population $p_a(x)=\mid\psi_a(x)\mid^2/(\mid\psi_m(x)\mid^2+\mid\psi_a(x)\mid^2)$ to a quadratic function of the propagation distance $x$, using the independently calibrated coupling constant $J_0$ as a frequency reference. Crucially, this procedure does not track any particle in time; it analyzes a static spatial interference pattern accumulated over many detection events. The resulting quantity $v$ has dimensions of speed, but it is a geometric parameter extracted from the spatial structure of the wave function, not a kinematical velocity obtained from the time evolution of particle positions.

With this conceptual clarification in place, we now demonstrate precisely what $v$ corresponds to within our weak actual value framework. In the evanescent regime, the wave function in the main waveguide is approximately

$$\psi_m(x)\propto e^{-\kappa x},\kappa=\frac{\sqrt{2m\mid\Delta\mid}}{\hbar},(24)$$

where $\kappa$ is the amplitude decay rate. This is a real-valued function in the stationary regime. The momentum weak value (14) for this wave function, with $\hat{p} = -i\hbar\,\partial_x$, is computed directly as

$$p_w = \frac{(\hat{p}\psi_m)(x)}{\psi_m(x)} = \frac{-i\hbar\,\partial_x e^{-\kappa x}}{e^{-\kappa x}} = i\hbar\kappa. (25)$$

This result follows immediately from our fundamental relation Eq. (14). The real part of the momentum weak value, $\text{Re}\,[p_w]$, is precisely the momentum weak actual value $p_w(t)$, which according to (18) equals the Bohmian particle momentum. For a purely evanescent wave with a real exponential form, we have

$$p_w(t) = \text{Re}\,[p_w] = \nabla S = 0, (26)$$

correctly indicating a vanishing Bohmian particle velocity, $v_B = p_w/m = 0$. The Bohmian particles are at rest in the steady state.

What about the parameter $\nu$? Comparing the experimental expression (23) with our result (25) establishes that

$$\frac{|\,\text{Im}\,[p_w]\,|}{m} = \frac{\hbar\kappa}{m} = \sqrt{\frac{2\,|\,\Delta\,|}{m}} = \nu. (27)$$

Thus, the experiment measures, up to a factor of mass, the magnitude of the imaginary part of the momentum weak value. This quantity is proportional to the spatial decay rate $\kappa$ of the wave function's amplitude. The decay rate is a geometric property of the guiding wave field, not a kinematical property of the Bohmian particle. The apparent paradox dissolves the moment one recognizes that the experiment simply does not operationally measure a particle speed; it probes the evanescent structure of the wave field, which is correctly described by standard quantum mechanics and fully compatible with the Bohmian particle dynamics.The steady-state particle distribution $|\psi(x)|^2$ observed in the detector is an ensemble effect. The time-integrated intensity converges to the stationary probability density. In the steady state, individual Bohmian particles are confined to the region $x \leq 0$ and have zero velocity; their distribution, however, extends exponentially into the forbidden region as dictated by the wave function. There is no conflict between stationary particles and an exponentially decaying wave field: this is precisely the Bohmian picture of a confined stationary state.

Proposed modifications to the guiding equation [15,17] that attempt to force Bohmian trajectories to match the geometric parameter $\nu$ are therefore misguided. They conflate wave function geometry with particle kinematics and abandon the principled ontological separation between wave and particle that is a central virtue of standard Bohmian mechanics. The standard guiding equation is uniquely determined by the requirements of Galilean covariance, time-reversal invariance, and the equivariance of the $|\psi|^2$ distribution [9]. Any ad hoc modification would violate these fundamental symmetries.

In summary, the experiment by Sharoglazova et al. provides an operational distinction between the real and imaginary parts of the momentum weak value, a distinction which our weak actual value framework makes theoretically precise. The real part (zero) correctly reflects the stationary Bohmian particles, while the imaginary part (nonzero) characterizes the exponential decay of the guiding wave. The alleged contradiction arises solely from a category error in the interpretation of the experimental data, which our analysis has definitively clarified.

## 6 Conclusion and outlook

We have presented a systematic analysis of Holland's local expectation values [8] within Bohmian mechanics, referring to them as weak actual values to emphasize their connection with weak measurement theory. As a conceptual starting point, we proposed and critically examined a heuristic criterion based on the local eigencondition (Remark 2.1), which serves to motivate the need for a more general theoretical tool. For general quantum superposition states, we studied this derived construct (Definition 3.1), demonstrating that its ensemble average equals the quantum expectation value (Property 1). Our primary contributions are threefold. First, we derived the exact time evolution equation for the weak actual value along a Bohmian trajectory (Theorem 2), extending Holland's primarily static analysis into the dynamical domain and providing a clear physical interpretation of the three contributing terms (Section 3.2). Second, we formally established the mathematical correspondence between the weak actual value and the real part of the weak value under position postselection (Eq. (15)), generalizing earlier insights to arbitrary observables. From this correspondence we extracted three concise lessons that articulate the physical connection between Bohmian mechanics and weak measurements (Section 3.3). Third, and most significantly, we deployed this unified framework to resolve a recent experimental challenge. Our analysis revealed that the measured quantity corresponds to the imaginary part of the momentum weak value, which reflects the spatial decay of the wave field, not the Bohmian particle velocity (Section 5). This cleanly separates particle kinematics from wave field geometry and rigorously eliminates the apparent contradiction. Illustrative applications to position, momentum, spin, and energy were given, demonstrating consistency with quantum measurement theory and foundational theorems such as Kochen–Specker and Bell.

Several limitations and open questions remain. First, the heuristic criterion is based on a local eigencondition; a fully rigorous justification would require a more sophisticated functional-analytic treatment. Second, the weak actual value and its evolution equation become singular at nodes of the wave function; under the quantum equilibrium hypothesis, such configurations form a set of measure zero, so statistical predictions remain unaffected, but a deeper understanding of the behavior near nodes is needed. Third, the generalization to quantum field theory presents significant challenges [10], particularly concerning infinite degrees of freedom, particle creation and annihilation, and renormalization.

Future research directions include: (i) establishing a rigorous functional-analytic foundation for the local eigencondition criterion; (ii) investigating the behavior of weak actual values near nodes using regularization techniques; (iii) extending the concept to Bohmian quantum field theory [10]; (iv) exploring relativistic and curved spacetime generalizations following Foo et al. [18,19]; (v) designing refined weak measurement experiments to test predictions based on the correspondence between weak actual values and weak values; (vi) investigating the behavior of weak actual values under non-equilibrium initial conditions for quantum thermodynamics and non-equilibrium quantum statistical mechanics.

Despite these limitations, this work provides a systematic and unified account of local observable properties in Bohmian mechanics, contributing to the dialogue between foundational theory and experiment, and offering a refined perspective on the physical interpretation of quantum phenomena.

## Acknowledgments

This research was supported by the National Natural Science Foundation of China (62475062,12547103), the Humboldt Research Fellowship Programme for Experienced Researchers (CHN-1218456-HFST-E), the Hainan Provincial Natural Science Foundation of China (124YXN412), and the Innovational Fund for Scientific and Technological Personnel of Hainan Province (KJRC2023B11).

## Data Availability Statement

No data were created or analyzed in this study.

## Conflict of Interest Statement

The author declares no conflict of interest.

## Appendix A: Consistency of the local eigencondition criterion

For completeness, we recall the proof of the equivariance theorem, which is the core statistical property of Bohmian mechanics. The probability density $\rho(\mathbf{q},t)=|\psi(\mathbf{q},t)|^2$ and the probability current density $\mathbf{j}(\mathbf{q},t)$ satisfy the continuity equation

$$\frac{\partial\rho}{\partial t}+\nabla\cdot\mathbf{j}=0, (A1)$$

where

$$\mathbf{j}_k(\mathbf{q},t) = \frac{\hbar}{m_k}\mathrm{Im}\,[\psi^*(\mathbf{q},t)\nabla_k\psi(\mathbf{q},t)].\,(A2)$$

The Bohmian velocity is $\mathbf{v}_k = \mathbf{j}_k/\rho$. A direct calculation of the material derivative of $\rho$ along a trajectory yields

$$\frac{d\rho}{dt} = \frac{\partial\rho}{\partial t} + \sum_k \mathbf{v}_k \cdot \nabla_k\rho = -\rho\sum_k \nabla_k \cdot \mathbf{v}_k.\,(A3)$$

Similarly, for $\ln\,|\,\psi\,|^2$, one obtains

$$\frac{d}{dt}\ln\,|\,\psi\,|^2 = -\sum_k \nabla_k \cdot \mathbf{v}_k.\,(A4)$$

Comparing (A3) and (A4) shows that $d(\ln\,\rho)/dt = d(\ln\,|\,\psi\,|^2)/dt$. Integrating along a trajectory and using the initial condition $\rho_0 = |\,\psi_0\,|^2$ gives $\rho(Q(t),t) = |\,\psi(Q(t),t)\,|^2$. By the quantum equilibrium hypothesis, this holds for all trajectories, establishing equivariance. □

# Appendix B: Detailed derivations of weak actual value properties

### B.1 Derivation of the evolution equation (Theorem 2)

We provide a detailed derivation for a single particle; the multi-particle generalization follows analogously. We assume that $\psi$ and $\hat{A}\psi$ are sufficiently smooth in a neighborhood of $(Q(t),t)$ to allow all pointwise operations.

Let $\psi(x,t)$ satisfy the Schrödinger equation

$$i\hbar\frac{\partial\psi}{\partial t} = \hat{H}\psi, \hat{H} = -\frac{\hbar^2}{2m}\frac{\partial^2}{\partial x^2} + V(x).\,(B1)$$

The complex conjugate equation is

$$-i\hbar\frac{\partial\psi^*}{\partial t} = (\hat{H}\psi)^*.\,(B2)$$

The Bohmian velocity is $v = J/\rho$ with $J = (\hbar/m)\mathrm{Im}\,(\psi^*\,\partial_x\psi)$ and $\rho = |\,\psi\,|^2$. The continuity equation reads

$$\frac{\partial\rho}{\partial t} + \frac{\partial}{\partial x}(\rho v) = 0.\,(B3)$$

Define $N(x,t) = \text{Re}\,[\psi^* \hat{A}\psi]$, so that $a_w = N/\rho$. The total time derivative along the trajectory is

$$\frac{da_w}{dt} = \frac{\partial a_w}{\partial t}\,|_{x=Q(t)} + v\frac{\partial a_w}{\partial x}\,|_{x=Q(t)}. (B4)$$

Using $a_w = N/\rho$,

$$\frac{\partial a_w}{\partial t} = \frac{1}{\rho}\frac{\partial N}{\partial t} - \frac{N}{\rho^2}\frac{\partial \rho}{\partial t}, \frac{\partial a_w}{\partial x} = \frac{1}{\rho}\frac{\partial N}{\partial x} - \frac{N}{\rho^2}\frac{\partial \rho}{\partial x}. (B5)$$

Substituting into (B4)

$$\frac{da_w}{dt} = \frac{1}{\rho}\left(\frac{\partial N}{\partial t} + v\frac{\partial N}{\partial x}\right) - \frac{N}{\rho^2}\left(\frac{\partial \rho}{\partial t} + v\frac{\partial \rho}{\partial x}\right). (B6)$$

Recognize the material derivatives

$$\frac{dN}{dt} = \frac{\partial N}{\partial t} + v\frac{\partial N}{\partial x}, \frac{d\rho}{dt} = \frac{\partial \rho}{\partial t} + v\frac{\partial \rho}{\partial x}. (B7)$$

From the continuity equation, $d\rho/dt = -\rho\,\partial v/\,\partial x$. Now, using the Schrödinger equation and noting that $\hat{A}$ is time-independent, so $\frac{\partial}{\partial t}(\hat{A}\psi) = \hat{A}\frac{\partial \psi}{\partial t}$, we have

$$\frac{\partial}{\partial t}(\psi^* \hat{A}\psi) = \frac{\partial \psi^*}{\partial t}\hat{A}\psi + \psi^* \hat{A}\frac{\partial \psi}{\partial t} = \frac{1}{i\hbar}\left[-(\hat{H}\psi)^* \hat{A}\psi + \psi^* \hat{A}\hat{H}\psi\right]. (B8)$$

Hence

$$\frac{dN}{dt} = \text{Re}\,\left[\frac{1}{i\hbar}\left(\psi^* \hat{A}\hat{H}\psi - (\hat{H}\psi)^* \hat{A}\psi\right) + v\frac{\partial}{\partial x}(\psi^* \hat{A}\psi)\right]. (B9)$$

Finally,

$$\frac{da_w}{dt} = \frac{1}{\rho}\text{Re}\,\left[\frac{1}{i\hbar}\left(\psi^* \hat{A}\hat{H}\psi - (\hat{H}\psi)^* \hat{A}\psi\right) + v\frac{\partial}{\partial x}(\psi^* \hat{A}\psi)\right] + a_w\frac{\partial v}{\partial x}. (B10)$$

Evaluating at $x = Q(t)$ yields equation (10) for the single-particle case. For multi-particle systems, the sum over particles appears analogously. □

### B.2 Specific form of the momentum weak actual value

For a single particle, write $\psi = Re^{iS/\hbar}$. Then

$$\hat{p}\psi = -i\hbar\nabla(Re^{iS/\hbar}) = e^{iS/\hbar}(-i\hbar\nabla R + R\nabla S). (B11)$$

Thus

$$\psi^* \hat{p} \psi = R e^{-iS/\hbar} e^{iS/\hbar} (-i\hbar \nabla R + R\nabla S) = -i\hbar R \nabla R + R^2 \nabla S. (B12)$$

Taking the real part, $-i\hbar R \nabla R$ is purely imaginary, so

$$\text{Re}\,[\psi^* \hat{p} \psi] = R^2 \nabla S. (B13)$$

Therefore

$$p_w = \frac{R^2 \nabla S}{R^2} = \nabla S = mv. (B14)$$

### B.3 Specific form of the energy weak actual value

For the Hamiltonian $\hat{H} = -\frac{\hbar^2}{2m}\nabla^2 + V$, acting on $\psi = R e^{iS/\hbar}$:

$$\nabla^2 \psi = e^{iS/\hbar} \left[ \nabla^2 R + \frac{2i}{\hbar} \nabla R \cdot \nabla S + \frac{i}{\hbar} R \nabla^2 S - \frac{1}{\hbar^2} R (\nabla S)^2 \right]. (B15)$$

Hence

$$\hat{H} \psi = e^{iS/\hbar} \left[ -\frac{\hbar^2}{2m} \nabla^2 R - \frac{i\hbar}{m} \nabla R \cdot \nabla S - \frac{i\hbar}{2m} R \nabla^2 S + \frac{1}{2m} R (\nabla S)^2 + VR \right]. (B16)$$

Multiplying by $\psi^* = R e^{-iS/\hbar}$ gives

$$\psi^* \hat{H} \psi = R \left[ -\frac{\hbar^2}{2m} \nabla^2 R - \frac{i\hbar}{m} \nabla R \cdot \nabla S - \frac{i\hbar}{2m} R \nabla^2 S + \frac{1}{2m} R (\nabla S)^2 + VR \right]. (B17)$$

Separating real and imaginary parts:

$$\begin{aligned} \text{Re}\,[\psi^* \hat{H} \psi] &= R \left[ -\frac{\hbar^2}{2m} \nabla^2 R + \frac{1}{2m} R (\nabla S)^2 + VR \right] \\ &= R^2 \left[ \frac{(\nabla S)^2}{2m} + V - \frac{\hbar^2}{2m} \frac{\nabla^2 R}{R} \right]. \end{aligned} (B18)$$

Thus

$$E_w = \frac{(\nabla S)^2}{2m} + V - \frac{\hbar^2}{2m} \frac{\nabla^2 R}{R} = \frac{(\nabla S)^2}{2m} + V + Q, (B19)$$

where $Q = -\frac{\hbar^2}{2m} \frac{\nabla^2 R}{R}$ is the quantum potential.